\documentstyle[aps,preprint,floats,epsfig]{revtex}
\begin{document}
\tighten

\title{Generalized Drell-Hearn-Gerasimov Sum Rule \\ 
  at Order ${\cal O}(p^4$) in Chiral Perturbation Theory}
\author{Xiangdong Ji, Chung-Wen Kao and Jonathan Osborne}
\bigskip

\address{
Department of Physics \\
University of Maryland \\
College Park, Maryland 20742 \\
{~}}

\date{UMD PP\#00-028 ~~~DOE/ER/40762-197~~~ October 1999}

\maketitle

\begin{abstract}
We calculate the forward spin-dependent 
photon-nucleon Compton amplitudes $S_1$ and $S_2$
as functions of photon energy $\nu$ and mass $Q^2$ 
at the next-to-leading 
(${\cal O}(p^4)$) order in chiral perturbation 
theory, from which we extract the contribution to 
a generalized Drell-Hearn-Gerasimov sum rule 
at low $Q^2$. The result indicates a much rapid 
$Q^2$ variation of the sum rule than 
a simple dimensional analysis would yield. 


\end{abstract}
\pacs{xxxxxx}

\narrowtext

The Drell-Hearn-Gerasimov (DHG) sum rule for the spin-1/2
nucleon relates its anomalous magnetic moment to an
integral over the spin-dependent photoproduction cross
section\cite{DHG}. Denoting the energy-dependent total 
photoproduction
cross sections with nucleon spin aligned or anti-aligned
with photon helicity as $\sigma_P(\nu)$ and $\sigma_A(\nu)$, 
respectively, one may write the DHG sum rule as
\begin{equation}
    \int^\infty_{\nu_{\rm in}}
      d\nu{\sigma_P(\nu)-\sigma_A(\nu)\over \nu}
    = {2\pi^2\alpha_{\rm em} \kappa^2\over M^2} \ ,
\label{DHG}
\end{equation}
where $\kappa$ and $M$ are the anomalous magnetic moment
(dimensionless) and mass of the nucleon, respectively, 
and $\nu_{\rm in}$ is the inelastic threshold. 
Because of the rapid development of the 
experimental technology in recent years, it now 
appears feasible to test the above sum rule which was derived
from Low's low-energy theorem \cite{low} and the assumption
of an unsubtracted dispersion relation \cite{DHG}.  

On the other hand, polarized electron-nucleon 
scattering provides a convenient way to measure the 
spin-dependent {\it virtual}-photon production
cross section on the nucleon. It is then interesting
to explore theoretically if there is a rule that 
governs the sum (or more precisely, the weighted 
integral) over the virtual-photon cross section. 
If such a sum rule exists, it clearly represents
a generalization of the Drell-Hearn-Gerasimov
sum rule to finite virtual photon mass $Q^2$. 

In a previous paper by two of us \cite{jiosborne}, 
we pointed out that because the 
DHG sum rule was derived from a dispersion relation 
for the invariant photon-nucleon Compton amplitude 
$S_1(\nu, Q^2=0)$, a generalized sum rule can be
naturally constructed from the same
dispersion relation at nonzero $Q^2$, 
\begin{equation}
   \int^\infty_{\nu_{\rm in}}
       G_1(Q^2, \nu){d\nu\over \nu}
    = {1\over 4}~\overline{S_1}(0,Q^2) \ .
\end{equation}
The above sum rule relates the integral of the nucleon 
spin-dependent structure function $G_1(\nu, Q^2)$ to 
the Compton amplitude $\overline{S_1}(\nu=0,Q^2)$, where
the overline denotes the subtraction of the contribution from
the elastic intermediate state. To endow 
the sum rule with physical content, one needs to 
find ways to calculate $\overline{S_1}(\nu=0, Q^2)$,  
extending Low's low-energy theorem. For the 
nucleon system at small $Q^2$, chiral perturbation 
theory ($\chi PT$) provides a natural tool. 
In Ref. \cite{jiosborne}, we found that $
\overline{S_1}(\nu=0,Q^2)$ is independent of 
$Q^2$ at ${\cal O}(p^3)$. In this letter, 
we report our result at next-to-leading order, 
${\cal O}(p^4)$.  

We first establish our notation and conventions. 
The forward photon-nucleon Compton scattering tensor
is   
\begin{equation}
     T^{\mu\nu} = i\int d^4\xi e^{iq\cdot \xi}
     \langle PS|{\rm T} J^{\mu}(\xi) J^{\nu}(0)|PS\rangle \ , 
\end{equation}
where $|PS\rangle$ is the covariantly-normalized
ground state of a nucleon with momentum $P^\mu$ 
and spin polarization $S^\mu$.
$J_\mu = \sum_i e_i \bar \psi_i\gamma_\mu \psi_i$
is the electromagnetic current (with $\psi_i$ being
the quark field 
of flavor $i$, and $e_i$ its charge in units of the proton charge).
The four-vector $q^\mu$ is the photon 
four-momentum. Using Lorentz symmetry, parity and 
time-reversal invariance, one can express the spin-dependent 
($\mu\nu$ antisymmetric) part of $T^{\mu\nu}$ in terms 
of two scalar functions:
\begin{equation}
     T^{[\mu\nu]}(P,q,S)  = -i\epsilon^{\mu\nu\alpha\beta}
      q_\alpha \left[S_\beta S_1(\nu, Q^2)
      +
    \left(\nu\, S_\beta -S\cdot q \,P_\beta/M\right) S_2(\nu, Q^2) \right] \ ,
\end{equation} 
where $\epsilon^{0123}=+1$ and $S_{1,2}(\nu, Q^2)$ are 
the spin-dependent, invariant Compton 
amplitudes. Through crossing symmetry, 
it is easy to see that $S_1(\nu,Q^2)$ is even in 
$\nu$ and $S_2(\nu,Q^2)$ is odd.

If we restrict ourselves to low $Q^2$ and $\nu$, 
$S_i(\nu,Q^2)$ are low-energy observables and 
hence it is natural to explore their physical content in 
chiral perturbation theory, or more broadly 
in low-energy effective theories. In $\chi$PT, one
regards the pion mass $m_\pi$ and the external 
three-momentum $\vec{p}$
small compared to any other scales in the problem.
In low-energy effective field theories one also 
considers expansions in terms of other small parameters, 
such as the mass difference $\Delta$
between the nucleon and delta resonance. The 
expansion parameter is then generically 
denoted as $\epsilon$. In this study, we will ignore
the delta resonance effects which seem to be small for 
$S_i(\nu, Q^2)$ \cite{jiosborne}. 

At order ${\cal O}(p^3)$, there is no contribution
to $\overline{S_1}(\nu=0,Q^2)$ \cite{jiosborne}. This conclusion
follows from a simple physical fact: the spin-dependent
effects on a nonrelativistic particle must vanish
as the mass of the particle goes to infinity. 
Since the nucleon mass $M$ is numerically comparable to
$4\pi f_\pi$, it is useful to organize $\chi$PT
by formally taking the
nucleon mass to infinity (heavy-baryon chiral
perturbation theory) \cite{heavybaryon}. From dimensional
analysis, the order ${\cal O}(p^3)$ contribution to 
$\overline{S_1}(\nu=0,Q^2)$ is independent 
of $M$ and hence must be zero. 

At next-to-leading order, or ${\cal O}(p^4)$, 
$\overline{S_1}(0, Q^2)$ is inversely proportional to 
$M$. To understand the type of Feynman diagrams that
we have to consider, it is useful to recall some of the standard
infrared power counting in effective field theories
as formulated in the heavy-baryon approach. The 
full lagrangian (including nucleon, delta, photon, and pion fields)
can be expanded :
\begin{equation}
    {\cal L} = {\cal L}^{(1)} + {\cal L}^{(2)} 
     + {\cal L}^{(3)} + {\cal L}^{(4)} + ... \ , 
\end{equation}
where ${\cal L}^{(n)}$ contains terms of order $\epsilon^n$,
with one power of $\epsilon$ assigned to each derivative, pion
mass, photon field, nucleon-delta mass difference, etc.  
The infrared power
of a Feynman diagram is generated from the vertices
and the propagators. For polarized Compton scattering,
Feynman diagrams start at ${\cal O}(\epsilon^3)$. 
For tree diagrams at this order, one needs to consider
vertices at ${\cal O}(\epsilon^n)$ with $n\le 3$. 
For one-loop diagrams, however, we need to consider 
vertices at ${\cal O}(\epsilon)$ only. In general,
at order ${\cal O}(\epsilon^n)$, one needs to consider
vertices at order ${\cal O}(\epsilon^{n-2\ell})$ for
diagrams of $\ell$-loops. According to the above,
we find 20 nonzero Feynman diagrams and their close 
relatives at ${\cal O}(p^4)$.  
These diagrams are shown in Fig. 1, where the cross
in each diagram represents an insertion from ${\cal L}^{(2)}$. 
 
\begin{figure}
\label{fig1}
\epsfig{figure=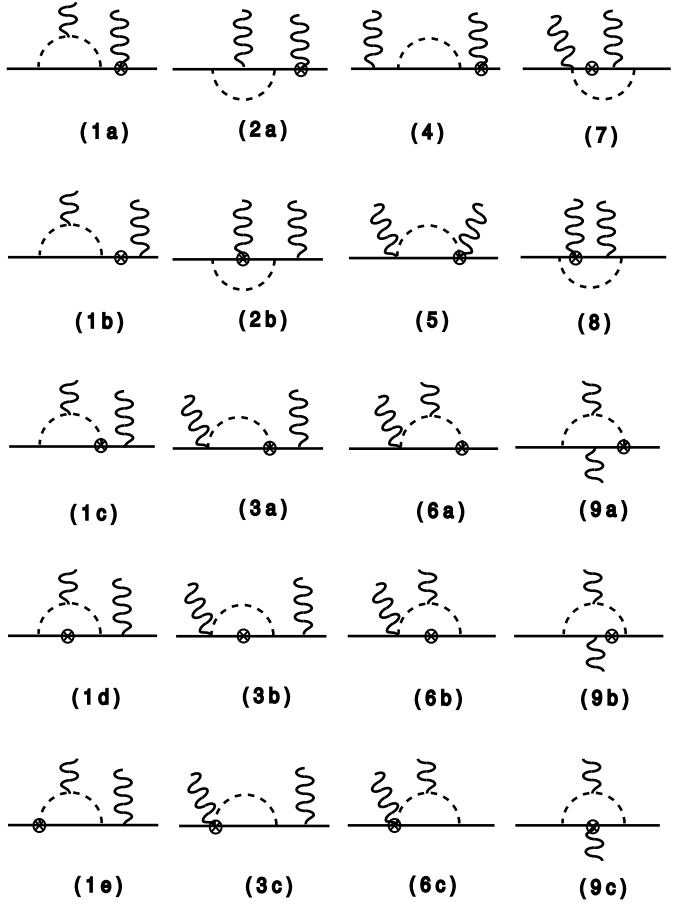,height=12cm}
\vspace{0.3in}
\caption{The diagrams that contribute to 
$S_{1,2}(\nu,Q^2)$ at NLO in heavy baryon $\chi$PT.  Obviously,
the diagrams from hermiticity and crossing must 
also be included.  The cross
indicates an insertion from ${\cal L}^{(2)}$.} 
\end{figure}   

We have calculated $S_1(\nu, Q^2)$ as a function of the photon
energy $\nu$ and mass $Q^2$. The result in the real photon limit
was reported in Ref. \cite{jikaoosborne}. Here we are interested
in $Q^2$ dependence at $\nu=0$, 
\begin{eqnarray}
     \overline {S_1}^{{\cal O}(p^4)}(0,Q^2) &=& 
{g_A^2\pi m_\pi\over 8(4\pi f_\pi)^2M} \left[
   -2(5+6\kappa_V+(1+6\kappa_S)\tau^3) \right.
   \nonumber \\
   && ~~\left.+
      \left(4\left(5+6\kappa_V+(1+6\kappa_S)\tau^3\right)+ {Q^2\over m_\pi^2}
    \left(3+6\kappa_V + (3+10\kappa_S)\tau^3
   \right)\right)\right.
\nonumber\\
    &&   ~~\left. \times\,\,\sqrt{m_\pi^2\over Q^2}\sin^{-1}
\sqrt{Q^2\over 4m_\pi^2+Q^2}\,\,\right], 
\end{eqnarray}
where 
$\kappa_S=-0.120 $ and $\kappa_V = 3.706$
are the experimental values of the isoscalar and isovector 
anomalous magnetic
momentum of the nucleon, respectively, and $\tau^3=\pm 1$ for the 
proton and the neutron, respectively. 
The above expression vanishes at $Q^2=0$
because we have subtracted away 
the chiral correction to the low-energy theorem \cite{low}.  
Its slope at $Q^2=0$ is 
\begin{equation}
      \left.{d\overline {S_1}(Q^2)\over dQ^2}\right|_{Q^2=0}
   = {g_A^2\pi\over 12(4\pi f_\pi)^2Mm_\pi}
     [(1+3\kappa_V + 2(1+3\kappa_S)\tau^3]. 
\end{equation}
Following the convention in the literature, we define
a dimensionless quantity $\Gamma(Q^2) \equiv (M^2/4)\overline{S_1}(Q^2)$ 
and write the generalized DHG sum rule as
\begin{equation}
     M^2\int^\infty_{\nu_{\rm in}} G_1(\nu, Q^2)
   = \Gamma(Q^2) \ .
\end{equation} 
Our prediction at small $Q^2$ can be expressed as
\begin{eqnarray}
    \Gamma^p(Q^2) &=& -{\kappa^2_p\over 4} + 6.85~ Q^2 ({\rm GeV}^2)
 + ... \nonumber \\
    \Gamma^n(Q^2) &=& -{\kappa^2_n\over 4} + 5.54~ Q^2 ({\rm GeV}^2)
+ ...\,\,\,\, ,
\end{eqnarray}
where we have used the low-energy theorem and the result
from leading-order chiral perturbation theory.
The  
$Q^2$ variation of the generalized DHG sum rule 
is large, and in fact is about a factor of $\pi^2$ larger 
than what one expects from naive power counting. 

For virtual-photon scattering, there is a second
sum rule which involves the spin-dependent
structure function $G_2$ and the Compton amplitude
$S_2$, 
\begin{equation}
    \int^{\infty}_{\nu_{\rm in}} {d\nu\over \nu^2}
    G_2(\nu, Q^2) = {1\over 4} \overline{S_2}^{(1)}(Q^2) \ , 
\end{equation} 
where ${\overline S_2}^{(1)}(Q^2)\equiv\partial(
 S_2(\nu,Q^2)-S_2^{\rm el}(\nu,Q^2))/\partial\nu\,
\big|_{\nu=0}$ is the slope of the (elastic) subtracted 
$S_2$ at $\nu=0$. In Ref. \cite{jiosborne}, we have 
obtained the chiral prediction for $\overline{S_2}^{(1)}
(Q^2)$ at leading order.
Now we have also the result for $\overline{S_2}^{(1)}(Q^2)$ 
at next-to-leading order, 
\begin{eqnarray}
      \overline {S_2}^{(1){\cal O}(p^4)}(Q^2) &=& 
{g_A^2\pi m_\pi\over 8(4\pi f_\pi)^2M} \left[
   -2\left(3(5+\tau^3)+8\left(\kappa_V-{Q^2\over m_\pi^2}\right)\right) \right.
   \nonumber \\
   && ~~\left.+ \left(12(5+\tau^3)+32\kappa_V
+{Q^2\over m_\pi^2}\left(9(1+\tau^3)
       +4(3\kappa_V+2\kappa_S\tau^3)\right)
          \right)\right.\nonumber\\
&& ~~\left. \times\,\,
\sqrt{m_\pi^2\over Q^2}\sin^{-1}\sqrt{Q^2\over 4m_\pi^2+Q^2}
\,\,\right].
 \end{eqnarray}
Numerically, this represents a large correction 
as one can see below.

   In Ref. \cite{bkm}, Bernard, Kaiser, and Meissner 
suggested to generalize the DHG sum rule using the 
combiniation ${\overline S_1}(Q^2)-Q^2{\overline S_2}^{(1)}
(Q^2)$. Because $\overline{S_1}(Q^2)=0$ at leading 
order in $\chi$PT, the $Q^2$ variation of their 
sum rule is related entirely to $\overline{S_2}^{(1)}(Q^2)$. 
Now, up to the next-to-leading order, we find 
\begin{eqnarray}
  &&  {\overline S_1}(Q^2)-Q^2{\overline S_2}(Q^2)= 
   -{\kappa^2\over M^2} + 
  8 {g_A^2\over (4\pi f_\pi)^2}\left(\sqrt{4m_\pi^2+Q^2\over Q^2}\tanh^{-1}
\sqrt{Q^2\over 4m_\pi^2+Q^2}-1\right) \nonumber\\
&& ~~-{g_A^2\pi m_\pi\over 4(4\pi f_\pi)^2 M}\left[-2\left(5+\tau^3+\kappa_V
-3\kappa_S\tau^3-4{Q^2\over m_\pi^2}\right)\right.\nonumber\\
  && ~~ +\left.\left(
4(5+\tau^3+\kappa_V-3\kappa_S\tau^3)+{Q^2\over m_\pi^2}(3+3\kappa_V+3\tau^3
-\kappa_S\tau^3)\right)\sqrt{m_\pi^2\over Q^2}\sin^{-1}\sqrt{Q^2\over
4m_\pi^2+Q^2}\,\,\right]\nonumber\\
&& ~~+{\cal O}(p^5) \ . 
\end{eqnarray}
The slope at small $Q^2$ is 
\begin{equation}
   \left. {d({\overline S_1}(Q^2)-Q^2{\overline S_2^{(1)}}
    (Q^2))\over dQ^2}\right|_{Q^2=0}
=\left({g_A\over 4\pi m_\pi f_\pi}\right)^2
   \left[{2\over 3}-{\pi\over 6}
 {m_\pi\over M} (13+2\tau^3+2\kappa_V)\right] . 
\end{equation}
The first term in the brackets was first obtained
in Ref. \cite{bkm} and, as we have stated, comes from the 
leading order $\overline{S_2}^{(1)}(Q^2)$. The subleading 
correction is our new result. Because of the large coefficient,
$13+2\kappa_V\sim 20.4$, the ${\cal O}(p^4)$ 
contribution is about a factor of 2 larger than the leading
contribution, and has the opposite sign.  As in the case 
of the spin polarizability \cite{jikaoosborne}, 
we question the validity of the chiral expansion 
of this quantity.

To conclude, we have calculated the ${\cal O}(p^4)$ 
correction to the generalized DHG sum rules. 
In a natural scheme introduced in Ref. \cite{jiosborne}, 
the $Q^2$ dependence of the generalized 
sum rule is much larger than
expected from a naive dimensional analysis. 
The same phenomenon is presumably responsible 
for an overwhelming next-to-leading-order 
correction to the rule considered by
Bernard, Kaiser, and Meissner \cite{bkm}. 

\acknowledgements
This work is supported in part by funds provided by the
U.S.  Department of Energy (D.O.E.) under cooperative agreement
DOE-FG02-93ER-40762.


\begin{references}
\frenchspacing

\bibitem{DHG}
S. D. Drell and A. C. Hearn, Phys. Rev. Lett. {\bf 16} 908 (1966);\\
S. B. Gerasimov, Sov. J. Nucl. Phys. {\bf 2} 430 (1966).

\bibitem{overview}
A. M. Bernstein and B. R. Holstein, Proceedings
on Chiral Dynamics: Theory and Experiment (Cambridge, MA, 1994). 

\bibitem{low}
F. Low, Phys. Rev. {\bf 96}, 1428 (1954); 
Phys. Rev. {\bf 110}, 974 (1958). 

\bibitem{jiosborne}
X. Ji and J. Osborne, hep-ph/9905410.

\bibitem{heavybaryon}
E. Jenkins and A. V. Manohar, Phys. Lett. {\bf B255} 558 (1991). 

\bibitem{jikaoosborne}
X. Ji, C.W. Kao, and J. Osborne, hep-ph/9908526.

\bibitem{bkm}
V. Bernard, N. Kaiser, and Ulf-G. Meissner, 
Phys. Rev. D {\bf 48}, 3062 (1993). 

 
\nonfrenchspacing
\end{references}
\end{document}